\documentclass[sigplan,screen]{acmart}
\AtBeginDocument{%
  \providecommand\BibTeX{{%
    \normalfont B\kern-0.5em{\scshape i\kern-0.25em b}\kern-0.8em\TeX}}}

\setcopyright{acmlicensed}
\copyrightyear{2018}
\acmYear{2018}
\acmDOI{XXXXXXX.XXXXXXX}

\acmConference[Conference acronym 'XX]{Make sure to enter the correct
  conference title from your rights confirmation emai}{June 03--05,
  2018}{Woodstock, NY}
\acmISBN{978-1-4503-XXXX-X/18/06}

\begin{document}

\title{Black Content Creators' Responses and Resistance Strategies on TikTok
}

\author{Gianna Williams}
\affiliation{%
  \institution{Northeastern University, Khoury College of Computer Sciences}
  \city{Boston}
  \state{MA}
  \country{USA}}
  \orcid{0009-0008-2363-3157}
\email{williams.gia@norhteastern.edu}

\renewcommand{\shortauthors}{Anonymous}

\begin{abstract}
   Social media wields a profound influence on social and economic dynamics worldwide \cite{10.1145/1242572.1242804}. With Content Creation, people on social media began to forge a livelihood through their online presence through creative labor.  This surge in social media Content Creators significantly shaped the trends and cultural landscape of the internet. While many of the social media trends we observe today can be attributed to the creative contributions of Black Content Creators \cite{doi:10.1080/08838151.2020.1815128}, digital platforms routinely marginalize and undermine these creators through algorithmic recommendation systems that produce systemic bias against Black and Brown people \cite{GuptaChirag2022RtIo}. To address this problem, we conducted a content analysis to assess how algorithms specifically illicit harassment, interact, and unfairly target Black Content Creators.  
\end{abstract}



\keywords{Do, Not, Us, This, Code, Put, the, Correct, Terms, for,
  Your, Paper}


\maketitle

\section{Introduction}

Social Media has grown over the past ten years. TikTok dates back to the early 2000s with MySpace to now 2023 with TikTok. Tiktok became popular in 2019 when Chinese ByteDance acquired the popular dance company muscia.ly . During the 2020 pandemic, Tiktok allowed people to quarantine to cope with COVID-19. Content Creators would make videos within various genres, including dance, makeup, cooking, comedy, etc. This app gave users a way to escape real-world tragedies.  
The term “content creator” rose in influence in 2021. A content creator can be “ someone who creates entertaining or educational material to be expressed through any medium or channel” \cite{Butler_2023}. With TikTok, content creators are promoted as people who can brand themselves for monetary gain. In 2020, Tiktok created the “Creator Fund”. The creator fund was implemented for eligible users with a  large following of 10k or above and 100k views per video. To some followers, this goal is easy, and to others, many have started to gamify their experiences by creating folk theories to develop a large following and gain more interactivity. \cite{10.1145/3465416.3483299}
Content creators are from many different geographic and demographic backgrounds and develop various types of content. As content creation has increased in popularity as a career, many creators rely on the obscure algorithms and regulations of video-sharing social media platforms. These opaque algorithms power these platforms and their followers' engagement to maintain relevance on the platform. More specifically, marginalized peoples have experienced inequities in the creator economy \cite{10.1145/3492841, 10.1145/3432951, article,Todd, Davis_2022,Pruitt-Young_2021}. These challenges include a heightened risk of experiencing harassment and aggressive content moderation \cite{10.1145/3491102.3517608,10.1145/3274424} unequal pay \cite{MSLStudyRevealsRacialPayGapinInfluencerMarketing_2021, Giacomazzo_2022}, and a lack of opportunities to grow their content. These issues are part of a more significant phenomenon of unequal treatment of marginalized groups by technological systems, from algorithm biases making unfair and inaccurate predictions and classification for marginalized groups \cite{10.1145/3442188.3445875, 10.1145/3465416.3483299, 10.1145/3531146.3533144}, to racism and bigotry on social media, causing these groups to experience heightened harassment, abuse, and mistreatment. Black content creators must work around these challenges to succeed in the creator economy.
Black Content Creators, in particular, have complained about their experiences on the TikTok platform. Researchers are starting to study this phenomenon through interview studies focused on content creators trying to make a living online. 
To explore how Black Content Creators use  \textit{Tik Tok} as a way to express themselves, we conducted a content analysis of the social media platform  \textit{Tik Tok} to investigate the creator's relationship with \textit{Tik Tok} content moderation. We saw that content creators had feelings of frustration that either led to community encouragement, leaving the platform, or resistance. 
We explored this through our research questions (RQs) below. 

\begin{itemize}
    \item (RQ1)How do social media platforms affect these Black Women Content Creators? 
\end{itemize}

\section{Related Work}

  Our paper plans to explore how social media platforms treat Black content creators. Before exploring our research, we believe it is imperative to understand how these social media platforms' algorithms are created, discuss how well they moderate content, and why Black Content Creators are essential to study within digital humanities and HCI. 
\subsection{Social Media Algorithms}
The mechanics that make these platforms run through content filtering algorithms. Content Filtered Algorithms “ recommend other items similar to what the user likes, based on their previous actions or explicit feedback” \cite{Akash._2023}. Figure 1 shows a diagram of the mental model used to define Content-based filter algorithms as they relate to targeted ads. 

\begin{figure}[htbp]
  \centering
  \includegraphics[width=0.8\linewidth]{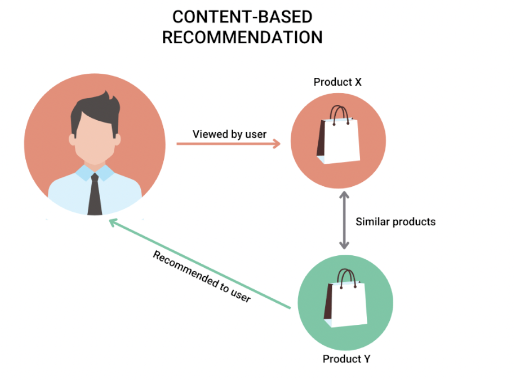} 
  \caption{Mental Model of Content-Based Filtering}
  \label{fig:example1}
\end{figure}

Regarding recommender systems in the context of the creator economy on social media platforms, research has explored how TikTok personalizes content for average users. \cite{10.1145/3485447.3512102, 10.1145/3604915.3608841,10.1145/3604915.3608825} 
As TikTok is a video social media platform, the app can become heavily saturated with content that might be flagged as harmful. Content moderation is used to monitor hate speech, 
 For example, Reddit banned its largest pro-Trump community, “r/The\_Donald”, after first quarantining the community \cite{Twitter_2021}. In January 2021, Twitter permanently suspended @realDonaldTrump account  \cite{Twitter_2021}, and Facebook subsequently suspended Trump’s accounts indefinitely \cite{Zuckerberg_2021} after attaching warning labels on violating posts. It seems that social media platforms are removing conservative speech, but examining the events that caused these moves reveals them as responses to onsight in real-life physical violence. Researchers have argued that social media platforms’ responses to defiance of their rules are to resist further violence and not censor users \cite{10.1145/3479610}.
This is a beneficial example of monitoring harmful content that can lead to physical violence offline. TikTok has been seen to disproportionately moderate marginalized communities' content, leading to fewer Black content creators on their platforms.  
\subsection{Shadowbanning}
Content moderation is used in social media platforms to regulate content. However, content moderation becomes more visible for users who are removed from the platform.
For instance, social media can also institute shadowbanning, suppressing the content creator's content until the moderator resolves their account. Social media like Instagram or TikTok bans hashtag discussions that may be polarizing \cite{10.1145/3479610}. This form of moderation affects content creators' mobility on the platform and may cause them to start over or leave the platform. 
Content creators being moderated have been significantly affected by future behaviors and have criticized these platforms for heavy censorship of their own free expression \cite{10.1145/3359294}. What makes content moderation different on platforms such as TikTok compared to other social media like Twitter or Facebook is that the algorithmic punishments have socioeconomic effects on these content creators. Tiktok’s creator fund is only accessible to those who receive 10k followers and above and 100k in viewership. If the platform continuously represses content, how can these creators earn a living? The demonetization adjusts content creators' behavior on the platform as they soon start to develop folk theories to resist shadowbanning. 
Ways content creators are resisting the platform shadowbanning is through AlgoSpeak language. HCI Researchers within the space of decoloniality have seen an increase in cultural hegemony of what is deemed valuable on social media platforms and what is not. Given how content moderation operates within the appeal process of being shadowbanned, Researchers see online communities creating their language to protect themselves from further restrictions. 
The earliest form of this can be seen within YouTube during the mid-2010s when Youtuber’s videos would be demonetized if they cursed or talked about sex. In 2023, TikTok creators are finding ways to get their content out and bypass excessive forms of moderation through algoSpeak. Examples of this include using emojis, replacing words, or mouthing words to reduce flagging. 
What is quite interesting is how these regulations specifically target marginalized content creators. Researchers have seen that Black content creators are known to get banned or removed from the platform for hate speech based on using African American Vernacular (AAVE)  that is specific to their own identities. Understanding the history of shadowbanning within the context of Black Content Creators will facilitate more conversation around the restrictive moderation that affects Black Content Creators' ability to grow on the platform. It is not that content creators are not on TikTok but that they are suppressed. 
\subsection{Black Content Creators }
André Brock's work on Twitter and other social media has created the stepping stone of Black digital studies within social media.\cite{alma9952023810901401} Black Twitter, for instance, became famous in 2009, \cite{Parham_2021} where Black Twitter users would live tweet Television, movies, and or music. This growth began the rise of live tweeting and streaming online for online communities. 
Limited work has been done on Black content creators' experiences on TikTok. However, there has been a growing interest in Black content creators' experiences within the CHI and CSCW. Previous work on exploring Blackness on social media focused on how users cultivate community. André Brock and Sarah Florini \cite {alma9952023810901401,doi:10.1080/08838151.2020.1815128} analyzed  Black Twitter.   Also adding on to research Catherine Mcknight Steele \cite{alma9952146751301401} as created digital form of Black Feminism:
\begin{itemize}
    \item \textbf{Intersectionality:} This principle, popularized by Kimberlé Crenshaw, emphasizes the interconnected nature of social categorizations like race, gender, class, sexuality, and more. Digital Black feminism recognizes how these intersections manifest in online spaces, affecting Black women uniquely due to their multiple marginalized identities. \cite{alma9952146751301401}
    \item \textbf{Centering Black Women's Experiences:} Digital Black feminism prioritizes Black women's experiences, voices, and perspectives in digital spaces. It seeks to address and rectify the erasure and marginalization they often face in online discourse. \cite{alma9952146751301401}
    \item \textbf{Tech Equity and Access:} It focuses on equitable access to technology and digital resources for Black women and marginalized communities. This includes advocating for improved digital literacy, fair representation in tech industries, and dismantling barriers to access.  \cite{alma9952146751301401}
    \item \textbf{Challenging Online Misogynoir:} Misogynoir refers to the specific intersection of racism and sexism directed towards Black women. Digital Black feminism seeks to confront and challenge online spaces where misogynoir is perpetuated, advocating for platforms to address harassment, discrimination, and hate speech targeting Black women.  \cite{alma9952146751301401}

    \item \textbf{Community Building and Solidarity: }Digital Black feminism promotes the creation of supportive online communities where Black women can connect, share experiences, and build solidarity. These spaces are essential for fostering empowerment, activism, and mutual support.  \cite{alma9952146751301401}
    \item Content Creation and Representation: Encouraging and promoting the creation of diverse and accurate representations of Black women in digital content is crucial. This involves challenging stereotypes, showcasing a variety of experiences, and fostering positive narratives.  \cite{alma9952146751301401}
    \item \textbf{Activism and Social Change}: Digital Black feminism utilizes online platforms for activism and advocacy, leveraging social media and digital tools to raise awareness about issues impacting Black women and to mobilize for social change.  \cite{alma9952146751301401}
\end{itemize}
Previous work finds ever-growing problems facing Black online communities, including racist harassment, infiltration of Black spaces, theft of Black cultural production, the commodity of Blackness, and excessive forms of Content Moderation. \cite{10.1145/3359294,10.1145/3479610,10.1145/3465416.3483299} These themes in the Black online communities help shape our work in performing a digital ethnography on specifically Black TikTokers.

\section{Methods}

This project explores the reality of Black Content Creator's experience on TikTok. This work specifically targeted Black creators on TikTok. This work aims to showcase what Black Women and non-binary creators face as they try to gain and remain relevant online. We also examine how Black creators’ perception of  \textit{Tik Tok} and what ways allow for growth. To do this, we performed a digital ethnography to explore and answer our research questions. 

\subsection{Demographics }

We ran our digital ethnography from October 2023 to November 2023. The content was obtained from the TikTok platform, where we saved all the content in a spreadsheet for qualitative coding. The content creators were 18 years old or above, with over 10k followers and 100k Views. We wanted to have these criteria based on what is described as a Content Creator on TikTok \textbackslash{}cite(Barnhart\_2023). The creators were from the Black Diaspora, with Black Women = 10 and Black Non-binary =1. In our findings, we labeled the accounts by A0X to differentiate the videos. 

\subsection{Ethnography Procedure}

We feed the TikTok search bar hashtags and queries to find the content. We searched and examined content until saturation.  

The hashtags searched were:

\begin{itemize}
    \item \#BlackGirlTikTok → This was for finding Black Women on TikTok
    \item \#BlackTikTok → This was for finding Black People on TikTok
    \item Black and TikTok  → This was a query that we used to find Black Content Creators
    \item \#TikTokGlitch → This was used later in the study as we found Creators expressing a “glitch” in the algorithm. 
\end{itemize}

\subsection{Content Analysis }

We organized the videos in a codebook with the following attributes: Video \# Video, Date Posted, Meta Data, Summary, Interpretation,  Theme, Demographics, Communication Style, Account Name, Follower Count, Type of Content Creator, Type Open Codes. Going further, we explain what and why we chose these attributes for our study. 

\begin{itemize}
    \item \textbf{Video \# : }To differentiate the type of videos from each other for analysis

    \item \textbf{Video Date Posted}:  We used this to observe the content from the creator better to see how their accounts progressed or regressed while on the platform.

    \item \textbf{Meta Data}: Metadata includes likes, captions, hashtags,  and view count. This is necessary to help understand what their audience felt about their content or if any agreed or disagreed.

    \item \textbf{Summary}: This was a visual summary of the content of the video

    \item \textbf{Interpretation}: Here, we interpreted the information for further examination of our study to help us create deductive codes

    \item \textbf{Demographics}: Here we noted the demographics of the content creator of either being a Black woman or a Nonbinary person

    \item \textbf{Account Name}: Here, we noted the account name to help us remember the account to help us differentiate the type of content creators.

    \item \textbf{Follower Count}: Here we noted the number of followers for the study to see if they are eligible or a part of the Content Creator community on TikTok

    \item \textbf{Type of Content Creator}: We noted the content creators based on their account audience to study how the algorithm showcased their content.

    \item \textbf{Open Codes}: We created this content to examine the qualitative data to capture the meaning and or aspects of the video in response to our research question.
\end{itemize}

Once we studied and coded the results, we performed a thematic analysis of the content to help us develop our findings. The researchers finalized a list of themes across multiple examinations to answer the research questions. Lastly, the finalized themes were categorized into greater overarching themes, and quotes from videos that supported each theme were selected.

\section{Findings}
Findings 
Below, we discuss the findings of digital ethnography by comparing them to our research questions. We first discuss our research question on how the algorithm affects Black Women Content Creators online. 

\subsection{Frustration}
	From our ten videos analyzed, we saw recurring themes of Black Content Creators expressing immense frustration on the platform. Many content creators we studied had a range in audience and following. We observed two comedy Tiktokers 2 Beauty Tiktokers 4 Misc (n=11) from our study. Content Creators would go to TikTok to discuss their experiences on the platform. Many of them would come from very intersectional identities. One Black, Non-Binary, Autistic Content creator discussed their frustration on why Black Autistic Content Creators are not being pushed on the platform. They go on to say that they were tagged by other users who were promoting Black Autistic Content Creators in a video. They then say, “Why have I never seen any of these accounts? And they have over 10k followers. So I feel like they should be on my For You Page”. (A05) This was a common theme from X amount of content creators studied where they saw lower video engagement. One comedy TikTok would say things like

 “The entire algorithm has changed… after a month of trying to figure out the algorithm and what works for me. Now I have to go back to the drawing board and figure out what TikTok wants to see… This honestly had me a little depressed a couple of days because I was like ‘wow’ I was almost there” (A02). This was a common theme found by those that were affected by the TikTok algorithm being optimized on trend. We saw some TikTok creators saying they worked aligned to what was popular on the platform. The folk theories used to regain popularity on the platform were TikTok language, such as “Girl Dinner,” or popular trending sounds. TikTok made it accessible to find what was trending by giving users access to trend sounds on the platform, as shown in Figure 2. Tren following was common among Black TikTokers who wanted to gain a following using sound from the  \textit{Tik Tok} Viral section. This finding leads us to our following finding: the shared feeling of community among the Black Content Creators. 

\begin{figure}[htbp]
  \centering
  \includegraphics[width=0.8\linewidth]{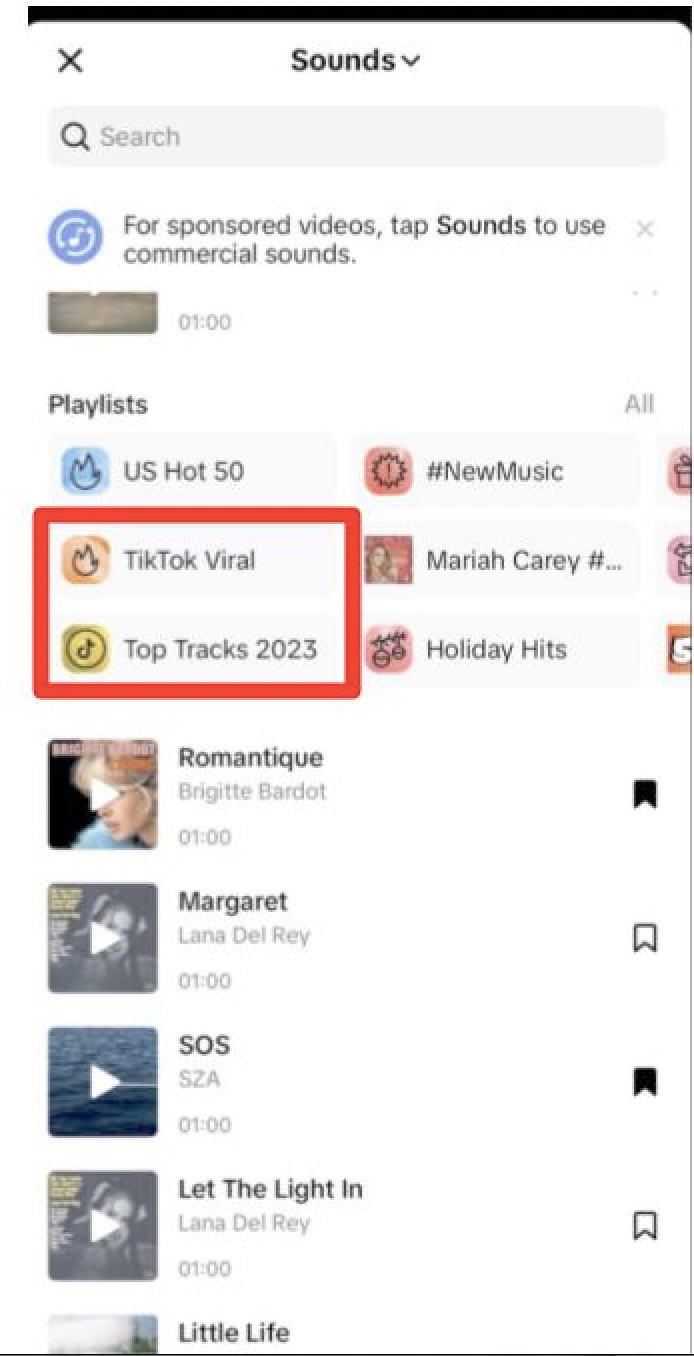} 
  \caption{TikTok Viral Feature}
  \label{fig:example2}
\end{figure}

\subsection{Community. }

From our study, we saw Content Creators not using the word “Shadow Banning” or Content Moderation but instead saying  “Tik Tok Glitch” (A02, A03). As these content creators received content suppression, many resisted these forms of invisibility online by advising users on how to stay relevant. A01 started her video by saying, “Anybody else experiencing a glitch on TikTok” (A01). The creator goes on to be positive by using affirming language, saying, “You are not alone if you are going through whatever is happening behind the scenes with this little glitchy, but we are going be back on track” (A01). This affirming language was also followed in the comments where users engaging with content expressed a similar concern. Saying things such as “I thought I was shadowbanned” (U01) and “Mine is acting a fool” U02. Some commenters offered advice based on what they were doing with the “Glitch” by keeping their content to themselves until the algorithm stopped suppressing it. This form of community was viewed in videos, comments, and hashtags.
Content Creators would use \#BlackgirlTiktok or \#BlackTikTok or \#BlackADHD to target their audience better. This folk theory is seen similarly within the formation of Black Twitters in the 2010s or shown through Pinterest before they created a more accessible way to find content based on Skintone. Those receiving content suppression expressed frustration; however, some never seemed to blame themselves or TikTok - rather, a simple “Glitch” in the algorithm. Those who became exhausted from their experience on TikTok stopped posting. One content creator gained immense traction during 2020 and 2021 and posted to TikTok in 2023 to explain her absence on the platform. Saying, “I think it is time for me to take an extended break. I do not know if or when I will return, and this is not me anymore… I am in such a different place in life... I have just grown up a little bit and I do not think this page reflects that growth” (A06). This creator explains that TikTok did not showcase her entire authentic self and did not want to be held to the person she was two years ago. 
\subsection{Authenticity }
Another common theme found within Content Creators was those who did not subscribe to specific sections of TikTok. We had 4 out of 11 accounts we studied that we labeled miscellaneous. What we saw from their content was creators pushing for authenticity. One miscellaneous content creator expressed why she left Instagram for content creation, and the reasoning was because she felt that “Back in  March my Instagram randomly got deleted, and I was like Oh My Gosh because not only was I making money over Instagram… it ended up being that I was using Instagram wrong… what I realized what is more important and what is more important long term is that you build a community. I think on Instagram, we focus so much on aesthetics.” A03.   She then explained that Instagram felt more like a job and not like her speaking to her friends. She also said that TikTok allowed her to show up to be more authentic. We also saw from other TikTok users the ability to pivot their content and not subscribe to what was trending on TikTok but more on what intrigued them. We will discuss this finding in our discussion regarding resistance and pleasure politics.  

\section{Discussion}
Our findings provide insight into the relationship between Black Content Creators and their interaction with the TikTok Platform. We add to a body of research that focuses on the experience of Black TikTokers \cite{10.1145/3610169, Pruitt-Young_2021}; we also give perspective on a community that finds ways to resist shadowbanning and content moderation of their platform. We apply the critical lens of Digital Black Feminist Thought to interpret Black content Creators' ways of selfhood and self-actualization in the face of suppressing and tightboxing their content. While this dynamic situates our work within a larger body of research on online communities in the CSCW and HCI communities, we propose a deeper investigation into the intersectional identities that choose to find joy within their content. 

\subsection{Encouragement of Black Content Creators Online}
In our study, we saw videos that, along with their content, creators would also post videos to affirm other Black Content Creators. Studies show marginalized communities are more likely to become shadowbanned for their content than others \cite{10.1145/3479610}. The content creators we studied relied a lot on their followers and others like them for support as they were undergoing homelessness. This account opened our eyes to the power of affirming, testifying, and community given from creators to creators. The act of testifying and affirming, as seen in Black Feminist Theory, involves sharing personal 

narratives to validate experiences and foster empowerment \cite{collins2022black, smitherman1986talkin}. Historically, This tradition allowed Black people to assert their realities and agency in a society that often silenced their voices.
Similarly, in gaming and online streaming, Black gamers are utilizing these platforms to share their gaming experiences, insights, and strategies. This sharing entertains, educates, and empowers others within the content creator community. This interaction helps build a community where stories, strategies, and experiences are shared and celebrated. Just as testifying in Black Feminist Theory served to build solidarity and create spaces for empowerment, through the act of vulnerability, online content creators foster a sense of community and belonging in the content creator world.

Along with testifying, those who suffered content moderation or the suppression of their content also connect with Black feminists' theory on community. Those in our study who had content suppression would create folk theories of how to regain visibility and use affirming language to ensure that the creators were not alone. This affirming language keeps content creators on the platform and allows them to find their audience. Account three, for example, had content starting in 2020 but gained more of a following from our study during the years of 2022 and 2023. After curating a community of followers online, she felt more comfortable showcasing her content online.

\subsection{Gamification of TikTokers}
In online content creation, Black Content Creators faced suppression, prompting many to engage actively with TikTok trends. One creator astutely remarked that her approach to TikTok was amiss, asserting that "all you need is a story, ring light, and good editing" (A08). Throughout our exploration, we consistently encountered a prevalent theme: content creators informing others about the latest trends favored by the algorithm.

Our paper contends with the gamification aspect inherent in TikTok, where creators strategically align themselves with patterns to maintain relevance. Accounts under scrutiny frequently adapted or diversified their content to appeal to varying target audiences, a tactic crucial to staying pertinent online.

Interestingly, one account ceased content production, attributing the decision to the belief that the algorithm confined her to a former self rather than embracing her present identity. This issue, echoing discussions prevalent in social media, notably surfaced earlier in the rise of YouTube. Extensive research exists on YouTubers departing from the platform due to their inability to keep pace with trends.

Our findings revealed a trend among creators: some conformed to TikTok's preferences rather than expressing their authentic content desires. This alignment with the platform's expectations often superseded their inclinations or intended content narratives.

Just as much as we saw Black Content Creators following trends to stay relevant. We also saw some content creators resisting TikTox trending optimization. 

\subsection{ Resistance of Content Creators}
In the dynamic landscape of social media, content creators often adapt their content to align with platforms' preferences, and TikTok is no exception. Our exploration revealed a prevalent trend among creators who strategically altered their content to cater to what they perceived TikTok's algorithm favored. This inclination to conform to the platform's expected trends and formats is a testament to the pervasive influence of algorithms in shaping content creation strategies.

Interestingly, our findings unveiled a compelling counter-narrative. Contrary to expectations, individuals whose content did not neatly fit into TikTok's expected mold did not necessarily voice dissatisfaction due to lower engagement. Surprisingly, many of these creators expressed a persistent desire to continue sharing content online despite needing to align more perfectly with a specific audience or trend.

This phenomenon resonates with "pleasure politics," where content creators consciously choose not to conform to popular trends or cast themselves into predefined molds. They opt not to project a constructed persona merely to appease algorithmic preferences. This defiance against conforming to the expected norms represents asserting individuality and resisting the pressure to conform to algorithmic dictates.

This deliberate choice not to tailor content to suit algorithmic expectations is an element of digital world-building. By prioritizing their authentic selves over algorithmic demands, content creators establish a space where self-actualization takes precedence. It is a realm where creators prioritize personal expression and authenticity, crafting their narrative worlds independent of the algorithm's influence.

This phenomenon points towards a shift where content creators reclaim agency and focus more on self-expression and authenticity rather than algorithmic appeasement. It signifies a shift towards a content creation landscape where individuality, personal narratives, and self-actualization supersede the pursuit of algorithmic success.

Understanding this divergence in content creation strategies sheds light on the multifaceted nature of online presence. It highlights the importance of recognizing and appreciating diverse motivations and content creators' strategies in navigating the digital sphere.

\section{Limitation and Future Work}

The study encounters several limitations that should be considered in interpreting the results. First and foremost, the sample size of the videos analyzed is limited. This constraint might restrict the breadth and depth of insights gained, potentially overlooking specific trends or perspectives in the broader TikTok landscape. Another significant limitation stems from selecting videos based on a single account. This approach may introduce bias due to individual preferences, demographics, or interactions.
Consequently, the findings might need to adequately represent the diversity of content available on TikTok, impacting the study's generalizability. Moreover, TikTok's algorithm-driven content distribution poses a challenge. The platform prioritizes trending content, potentially skewing the analyzed videos towards popular or trending topics rather than reflecting a comprehensive spectrum of user-generated content.

Future research should consider several avenues to address these limitations and advance the study. Leveraging the TikTok Research API could substantially enhance the dataset's size and diversity, allowing for a more comprehensive analysis of content trends and user behaviors. Additionally, diversifying the source of analyzed videos across multiple accounts representing various demographics or regions would mitigate individual bias, providing a more holistic understanding of TikTok's content landscape. Understanding the nuances of TikTok's algorithmic influence demands further investigation. Exploring the specific criteria that drive content visibility and popularity within the platform's algorithm would offer valuable insights into video dissemination and user engagement. Longitudinal studies tracking content trends and shifts would provide a dynamic perspective on evolving content preferences and platform dynamics. This longitudinal approach could uncover nuanced changes in user-generated content and audience engagement patterns. Finally, exploring alternative metrics beyond trend-based relevance, such as user engagement, diversity of topics, or user preferences, would refine TikTok's understanding of content relevance and enrich the analysis of video content within the platform.

In addressing these considerations, future research endeavors can mitigate existing limitations and offer a more comprehensive and nuanced understanding of TikTok user experiences and content trends.

\section{Conclusion}
Our research involved a comprehensive content analysis of TikTok to gain deeper insights into the experiences of Black Content Creators within the platform. This digital ethnography was conducted from October to November 2023, employing open coding techniques to analyze the amassed data meticulously.

Through our rigorous examination, we saw three prominent themes that encapsulated the essence of these creators' experiences: Frustration, Community Support for Content Creators, and the presence of self-actualized content creators. Delving deeper into our findings, we observed a correlation with the tenets of Black Feminist Theory. Specifically, the sense of community and testifying emerged as pivotal elements fostering support among content creators, encouraging their continued engagement with the platform.

Furthermore, our exploration unveiled a fascinating trend indicative of a gamified approach adopted by content creators. Many appeared to align their content creation strategies with the algorithm's preferences, prioritizing content that resonated with its anticipated priorities rather than their inclinations or narratives.

Lastly, our discussion illuminated the resistance encountered by sure creators who consciously chose not to conform to specific genres or prescribed content types prevalent on the platform. This resistance was notably prevalent among miscellaneous content creators, indicating a divergence from conforming to popular trends or predefined categories.

By uncovering these multifaceted dynamics, our study sheds light on the complexities inherent in the experiences of Black Content Creators on TikTok. It underscores the influence of algorithmic preferences and highlights the resilience and varied strategies creators employ to navigate and assert their presence within the platform's digital landscape.


\end{document}